\documentclass[sigconf]{acmart}

\begin{document}

\title{ON HOMOMORPHIC ENCRYPTION BASED STRATEGIES FOR CLASS IMBALANCE IN FEDERATED LEARNING}

\author{Arpit Guleria\\ J. Harshan}
\affiliation{%
  \institution{Department of Electrical Engineering\\
  IIT Delhi, New Delhi}
  \country{India}
}

\author{Ranjitha Prasad}
\affiliation{%
  \institution{Department of Electronics and Communications Engineering\\
  IIIT Delhi, New Delhi}
  \country{India}
}

\author{B. N. Bharath}
\affiliation{%
  \institution{Department of Electrical, Electronics and Communication Engineering\\
  IIT Dharwad, Karnataka}
  \country{India}
}



\begin{abstract}
  Class imbalance in training datasets can lead to bias and poor generalization in machine learning models. While pre-processing of training datasets can efficiently address both these issues in centralized learning environments, it is challenging to detect and address these issues in a distributed learning environment such as federated learning. In this paper, we propose FLICKER, a privacy preserving framework to address issues related to global class imbalance in federated learning. At the heart of our contribution lies the popular CKKS homomorphic encryption scheme, which is used by the clients to privately share their data attributes, and subsequently balance their datasets before implementing the FL scheme. Extensive experimental results show that our proposed method significantly improves the FL accuracy numbers when used along with popular datasets and relevant baselines.
\end{abstract}



\maketitle

\section{Introduction }
Federated Learning (FL) is a distributed privacy-preserving machine learning (ML) paradigm, where a global model is obtained by  aggregating models trained locally at each decentralized devices \cite{FLApplSurvey, Fedavg}. Advancements in the FL schemes with differential privacy and homomorphic encryption schemes make it  suitable for the scenarios where data privacy is of utmost concern \cite{FL_Homomorphic}. Although vanilla FL handles data privacy to a certain extent, it is not designed to handle issues like class imbalance \cite{Glo_Imbal} where the data set has skewed class proportions at the participating devices \cite{LearnImbData}. Moreover, sharing information on the skewed class proportions with the central entity may potentially hamper data privacy of the participating devices. Motivated by this, we propose novel methods to detect and address \textit{class imbalance} issues in FL environments without compromising on privacy.

\begin{figure}[htpb]
  \begin{center}
    \includegraphics[scale = 0.5]{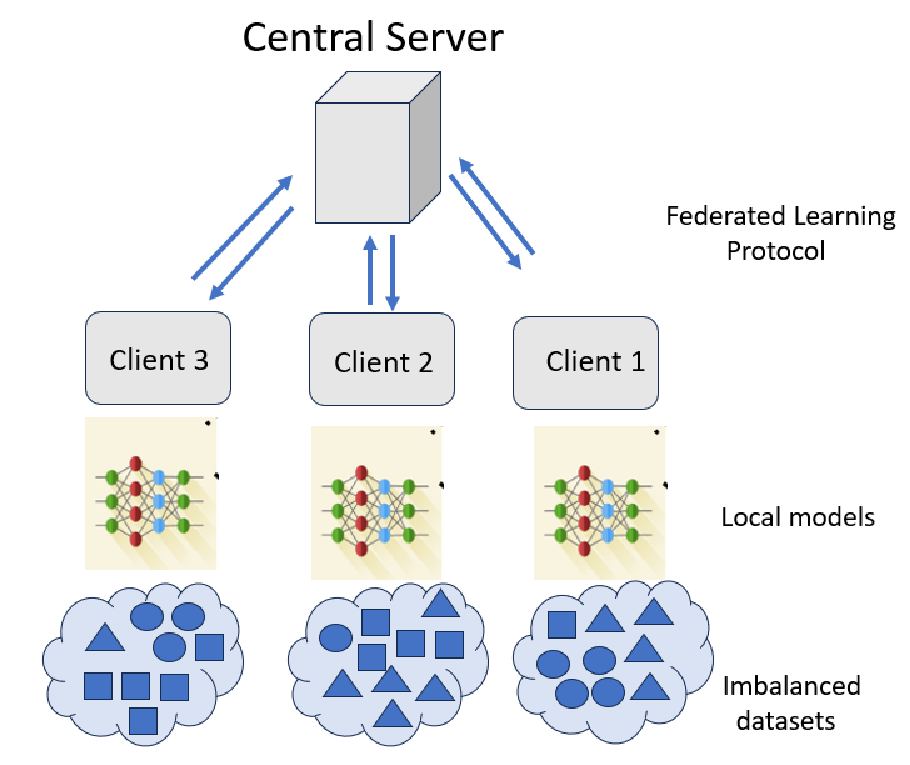}
    \vspace{-0.4cm}
    \caption{FL setup with $3$ clients trying to solve $3$-class (circle, triangle and square) problem with class imbalance.}
    \label{FL_big_pic}
  \end{center}
\end{figure}

A typical FL setup with class imbalance is shown in Fig.~\ref{FL_big_pic}. We refer to classes that make up a large and small proportion of the dataset as majority and  minority classes, respectively. Class imbalance has negative repercussions on the training of a classification model, as it leads to biased predictions and poor generalization. In centralized learning environments, class imbalance can be handled by using data level and algorithm level approaches. In the data level approach, pre-processing of training data is done by resampling methods or data augmentation methods \cite{datasamp}. In algorithm level, measures like Cost-Sensitive Learning, Ensemble Methods and Adaptive Synthetic Sampling (ADASYN) are used to deal with data augmentation \cite{ADASYN}. Detecting and addressing class imbalance in centralized learning environment is well-studied as compared to the FL environments. 

\subsection{Need for Global Imbalance Detection}

Consider the following toy example of an FL setup wherein two clients, namely client 1 and client 2 intend to interact with a server to generalize their models for a classification task. The objective is to train their neural network to classify unseen samples into one of the two classes, labelled \emph{Class 0} and \emph{Class 1}. Suppose that client 1 has 10 and 90 training data samples from \emph{Class 0} and \emph{Class 1}, respectively. Similarly, suppose client 2 has 90 and 10 training data samples from \emph{Class 0} and \emph{Class 1}, respectively. Evidently, if the clients communicate their data distribution to the server, they need not apply any correction to their local distributions since the global distribution of the data samples is perfectly balanced. However, if the two clients do not communicate their data distributions due to data confidentiality, they tend to locally apply data-level approaches of pre-processing their training data which was not necessary in an FL framework. Thus, this example shows that clients need to check with the global imbalance of their datasets before entering into FL process. 

\subsection{Need for Global Imbalance Correction}
Towards improving the class imbalance, assume that the two clients choose to locally pre-process their training samples either by down-sampling the majority class or over sampling the minority class.
If they choose to down-sample the majority class, then they may remove redundant data samples within their dataset. However, given the FL nature of collaboration, the features in the removed data samples might still be of interest to the other client, and this may lead to accuracy losses. On the other hand, if the clients choose to individually over-sample the minority classes through augmentation methods, then while the local imbalance may improve, the local datasets may have significant number of "artificial" data samples that are not close to reality. We refer to this fraction of data samples with respect to the original data samples as the augmentation-overhead. Clearly, a large augmentation-overhead is an impediment in internet of things-type systems that have memory constraints \cite{IOTSurvey}. Thus, it is desired that over-sampling is done in a prudent manner so as to reduce the augmentation-overhead in such systems. For the above mentioned reasons, we believe that the clients in the FL process, should collaborate to implement the resampling methods by tracking the improvements in the global imbalance metrics at the server. 

\subsection{Contributions}
Tracking global imbalance metrics may often require the clients to share sensitive information such as the distribution of the data or the data itself leading to privacy and security breach. Therefore, there is a need to compute and transmit information in a secure and privacy preserving manner. The choice of encryption is critical since it should not only keep the data secure but also should enable efficient class imbalance detection using which subsequent corrections can be made to the data. Identifying a promising research direction in this space, we  propose a novel strategy to detect and address global class imbalance in FL environments using a data level approach. For providing privacy to the clients' data distribution, we employ the Cheon-Kim-Kim-Song (CKKS) homomorphic scheme \cite{HEAN} wherein a global imbalance detection and corrections are implemented thereby improving the effectiveness of their FL solutions. Our specific contributions are as follows:
\begin{itemize}
    \item We propose \emph{FLICKER}, an iterative mechanism to detect and correct global imbalance among the datasets at the clients in an FL setting. It comprises a privacy-preserving step for secure computation of global imbalance at the server, a localization step to identify a client that has a dominant contribution to the global imbalance followed by an imbalance correction step at such a dominant client. With every round of imbalance correction at the dominant client, an update step on the global imbalance is executed, and this procedure is repeated until a target global imbalance is achieved.
    \item In order to compare a wide range of imbalance correction frameworks in FL, we propose a novel metric, denoted as the normalised accuracy, which refines the definition of the traditional accuracy metric by accounting for the augmentation-overhead, contributed by the over-sampling techniques during imbalance correction.
\end{itemize}
To showcase the benefits of FLICKER, we perform extensive experiments consisting of baseline comparisons and ablation studies on the image based CIFAR-10 and Natural Language Processing (NLP) based AG News dataset. As baselines, we consider two strategies wherein the clients apply over-sampling and under-sampling techniques on their data samples to correct their local imbalance without checking the imbalance in the global distribution. Our results demonstrate that the proposed interactive FLICKER outperforms the baselines in terms of the normalised accuracy and F1-scores of individual classes.

\subsection{Related Works and Novelty}
\label{PW}
In this section, we discuss relevant literature on FL related to class-imbalance issues and homomorphic encryption methods in FL. We also discuss the novelty of the proposed method. 


\textbf{Federated Learning and Class Imbalance}: Several techniques have been proposed to handle class imbalance in FL. These include client selection \cite{CIClientSelec}, data augmentation \cite{surveyimbalance}, adaptive aggregation \cite{Ijcai23} and federated reweighting. Class-imbalance in long-tailed distributions have been addressed in \cite{longtailed}. In \cite{CIF1}, \cite{CIF2} and  \cite{CIF3}, authors propose ways to address class imbalance in FL environment using algorithm level approaches. Algorithm level approaches are model specific and hence, not relevant in model agnostic scenarios. However, none of the above methods ensure a secure mode of communicating the class-imbalance to the server.

\textbf{Homomorphic Encryption and their applications in FL:} Homomorphic encryption methods are known to facilitate secure computations on the cipher-texts thereby enabling untrusted entities to perform algebraic operations on their plain-texts by preserving privacy. While the traditional fully homomorphic encryption methods are computation-intensive, recent class of homomorphic schemes for arithmetic of approximate numbers \cite{HEAN} has gained attention for providing flexibility in computational-complexity and accuracy. This scheme, which is also known as the CKKS encryption scheme, is based on lattice based cryptography wherein the concept of \textit{ring learning with errors} is used for \textit{implementing closest vector problem}. Due to the benefits offered by the CKKS scheme, they have found applications in securing FL \cite{CKKS_FL_1}, \cite{CKKS_FL_2}, \cite{CKKS_FL_3}, \cite{CKKS_FL_4}. Besides the CKKS specific contributions to FL, secure FL frameworks through generic homomorphic encryption schemes have also been addressed in the past \cite{FL_Homomorphic}, \cite{FE_FL_1}. \\
\textbf{Novelty}: To the best of authors' knowledge, this is the first instance where CKKS encryption is employed to identify the class imbalance issue in FL without compromising on privacy. At the heart of our strategy lies the incorporation of the cosine similarity between global and the local distribution in order to identify the clients that introduce imbalance.  Since cosine similarity computation is executed through homomorphic encryption via primitive addition and multiplication operations, this process of identification of the clients is done in a privacy-preserving manner. Furthermore, appropriate resampling methods are chosen at the clients in order to address the problem of global imbalance. We highlight that the server update process that reflects the balanced local distribution also preserves privacy as it is carried out using the CKKS homomorphic encryption method.

\section{FLICKER: CKKS Based Strategy for Global Imbalance Correction}

Consider an FL use case, wherein $N$ clients intend to collaborate to refine their ML models in order to perform a classification task on their dataset into one of the $L$ classes. Towards training their models, each client is equipped with training datasets of certain size. Let the sample distribution at the $n$-th client be $LD_{n} = [l_{n,1}, l_{n,2}, \ldots, l_{n, L}]$, where  $l_{n, j} \geq 0$ represents the number of training data samples from class $j$, for $1 \leq j \leq L$. As per \cite{CIF1}, the local balance metric at the $n$-th client, for $1\leq n \leq N$, is
\begin{equation}
\label{4.1}
  LI_{n} = \frac{min(LD_{n})}{max(LD_{n})}. 
\end{equation}
Based on \eqref{4.1}, a distribution can be referred to as a balanced distribution  if its balance metric is close to one. Otherwise, it is said to suffer from severe imbalance if \eqref{4.1} is close to zero. Typically, in FL, its test accuracy depends on the global distribution of the samples instead of the local distributions. In this context, the global distribution of training samples is $GD = \left[GD_1, GD_2, GD_3, \ldots, GD_{L}\right],$ 
where $GD_j = \sum_{n = 1}^{N}LD_{n,j}.$ Consequently, the balance metric on the global distribution is
\begin{equation}
\label{4.4}
GI = \frac{min(GD)}{max(GD)}. 
\end{equation}
Given that the model parameters of each client are shared with the other clients in FL, $GD$ needs to be estimated before proceeding with the FL algorithm. Typically, depending on the underlying use case, a minimum global imbalance of a certain threshold, namely $0 < \Delta < 1$, is treated as an acceptable value. However, when this threshold is violated, the clients may need to apply appropriate under-sampling or over-sampling methods on their datasets in order to pull this metric above the threshold.

In our model, we assume that each client imposes privacy requirement on their dataset, their local distribution, as  well as on its balance metric. Thus, while each client has its local distribution and the local balance metric, the server cannot have this information in plain-text in order to compute the global imbalance. Therefore, in the next section, we introduce FLICKER, a CKKS based homomorphic EncRyption strategy for data Imbalance Correction in Federated Learning.

\subsection{Methodology} \label{sec:methodology}

Our methodology, which relies on an iterative mechanism to detect and correct global imbalance, comprises four steps, namely: (i) A privacy-preserving step for secure global imbalance computation at the server, (ii) A localization step to identify a dominant client for imbalance correction, (iii) a local imbalance correction step at the dominant client, and finally, (iv) an update step on computing global imbalance. We assume that each client has its local distribution, i.e., $LD_{n}$, along with its local imbalance value, i.e., $LI_{n}$. The server implements a CKKS based homomorphic encryption method, wherein its public key is known to all the clients, whereas the private key is with the server to recover the computations on the ciphertexts. Therefore, CKKS encryption can be done by either the clients or the server. However, the CKKS decryption can be only be done by the server. Henceforth, the public key of the server for CKSS encryption is denoted by $PK_{ser}$. We also assume that all the clients can privately communicate among themselves in a pair-wise manner.

\begin{figure}[htpb]
  \begin{center}
    \includegraphics[scale = 0.5]{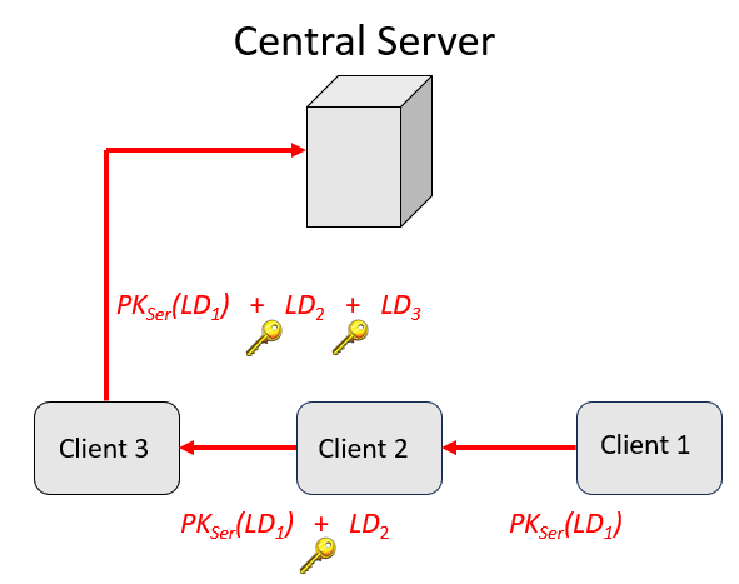}
    \vspace{-0.4cm}
    \caption{Depiction of distributed computation of global distribution at the central server through homomorphic message-passing in the three clients model. Keys adjacent to the addition operator indicates homomorphic addition.}
    \label{FL_global_imb}
  \end{center}
\end{figure}

\subsubsection{Secure Global Imbalance Computation}
\label{sec:sgic}

In this step, the clients form a virtual ring-network among them such that they share their local distribution to their neighboring clients, and forward a cumulative sum of their distributions in a privacy preserving manner using the CKKS scheme. Without loss of generality, we assume that the messages are passed in the order: client $1$ $\rightarrow$ client $2$ $\rightarrow$ $\ldots$ $\rightarrow$ client $N$ before reaching the server. At the first stage, client $1$ encrypts its local distribution $LD_1$ with the server's public key for CKKS operation, and sends ${PK_{ser}(LD_1)}$ to client $2$. Subsequently, client 2, after recovering the message from client $1$, forwards the message $PK_{ser}(LD_1) + LD_2$ to client $3$, which refers to the encrypted version of the sum of $LD_1$ and $LD_2$. By repeating similar operations at the other clients along the similar lines, client $N$ shares the following message with the sever 
\begin{equation}
\label{eq:homo_enc}
E_{PK_{ser}}[GD] \triangleq \sum_{n = 2}^{N} LD_{n} + PK_{ser}(LD_1).
\end{equation} 
Finally, using the CKKS scheme, the server uses its private key on $E_{PK_{ser}}[GD]$ to recover 
$GD = \sum_{n = 1}^{N} LD_{n}$ from the encrypted message in \eqref{eq:homo_enc} without learning the local distributions of the clients. This distributed method of computing the global distribution is exemplified in Fig. \ref{FL_global_imb} for an FL setup with $N = 3$.

Using the global distribution $GD$, the server thereafter calculates the global imbalance using \eqref{4.4}. If the measured $GI$ is above a preset threshold, say $\nabla$, for some $0 < \nabla < 1$, then the server and the clients initiate the FL algorithm using their datasets. On the other hand, if the measured $GI$ is below $\nabla$, the server detects a client contributing to this global imbalance.

\begin{figure}[htpb]
  \begin{center}
    \includegraphics[scale = 0.52]{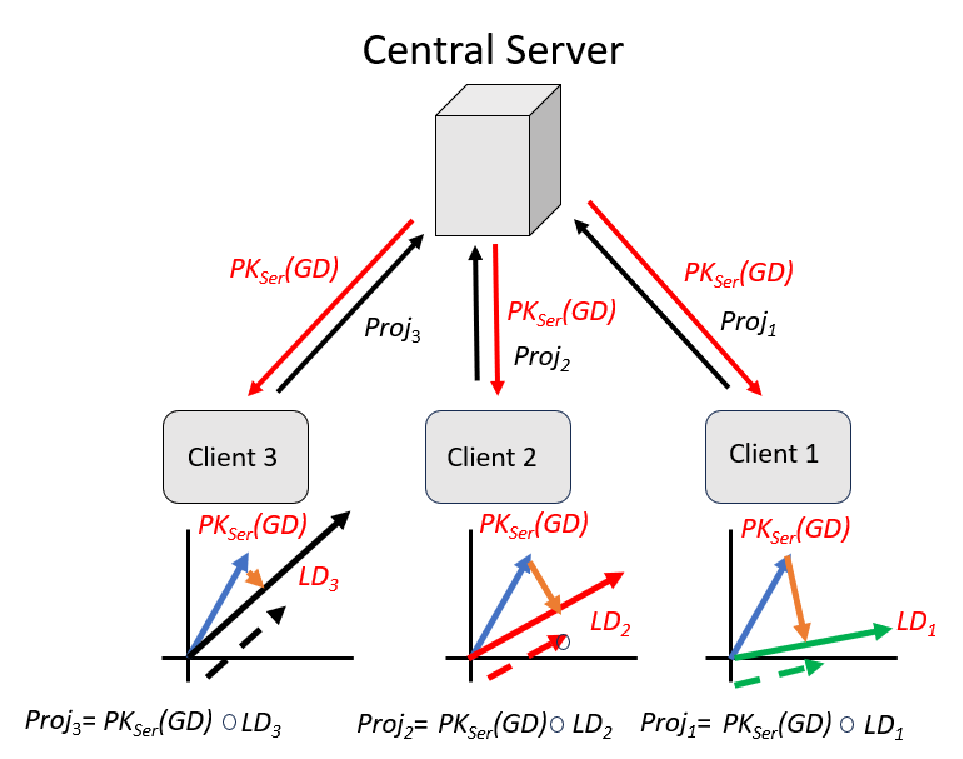}
    \vspace{-0.7cm}
    \caption{Depiction of steps to localize the dominant client through projection operation in the three clients model.}
    \label{dominant_client}
  \end{center}
\end{figure}

\subsubsection{Client Localization}
\label{sec:client_loc}

In this step, the server normalizes the global distribution vector to have unit norm, to obtain $GD_{norm} = \frac{GD}{||GD||},$ where $||GD||$ denotes the Frobenious norm by treating $GD$ as an $L$-dimensional real vector. Subsequently, it broadcasts $PK_{ser}(GD_{norm})$ to all the clients.

Upon receiving $PK_{ser}(GD_{norm})$, client $n$, for $1 \leq n \leq N$,  computes the cosine similarity between its local distribution and the encrypted version of $GD_{norm}$, as $CS^{E}_{n} = \frac{LD_{n}}{||LD_{n}||} \circ PK_{ser}(GD_{norm}),$ and returns this $CS^{E}_{n}$ to the server. Due to CKKS, $CS^{E}_{n}$ denotes the encrypted version of the cosine similarity value from client $n$. Here, the operator $\circ$ denotes the inner-product operation between two $L$-dimensional vectors. This method of localizing the dominant client is depicted in Fig. \ref{dominant_client} for an FL setup with $N = 3$. Given that the cosine similarity is computed in the cipher domain, the values returned by the clients also maintain the confidentiality feature. After recovering the plain-text version of the cosine similarities, the server obtains $CS^{P}_{n} = \frac{LD_{n}}{||LD_{n}||} \circ GD_{norm},$
for each client. Finally, the server picks the client that has maximum cosine similarity with the global distribution, and asks the corresponding client to correct its local imbalance by applying off-the-shelf over-sampling and down-sampling methods on its local datasets. Henceforth, the chosen client is referred to as the dominant client. 


\subsubsection{Resampling Procedure at the Dominant Client}

Our proposed method is such that the dominant client checks whether the local imbalance of its dataset is below a threshold, denoted by $\nabla_{client}$, for some $0 < \nabla_{client} < 1$, to apply a resampling procedure. Furthermore, if the local balance is less than $\nabla_{client}$, then the resampling procedure is implemented. Thus, for a given $\nabla_{client}$, the dominant client updates its local distribution at its end, and follows the procedure in the next section to update the global distribution at the server. 

\begin{figure}[htpb]
  \begin{center}
    \includegraphics[scale = 0.52]{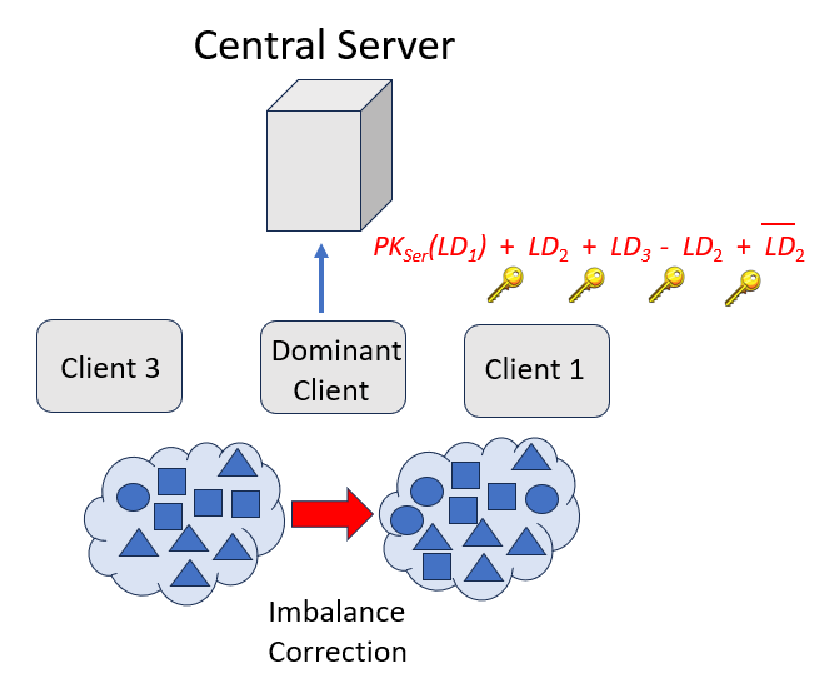}
    \vspace{-0.4cm}
    \caption{Depiction of local imbalance correction and subsequent update operation on global distribution in the three clients model. In this context $\overline{\mathbf{LD}_{2}}$ denotes the updated local distribution at the dominant client.}
    \label{global_update}
  \end{center}
\end{figure}

\subsubsection{Update Step on Global Imbalance}

In order to securely convey the update on local distribution, the dominant client subtracts its past local distribution from \eqref{eq:homo_enc}, and then adds the updated local distribution to the same compute before sending it to the server directly. This method of updating the global distribution after the imbalance correction at the dominant client is depicted in Fig. \ref{global_update}. Along with the update, the dominant also mentions whether it has achieved the maximum allowed local imbalance through the current round. Since the update is made on the encrypted cipher, the CKSS decrryption algorithm at the server ensures that the new global distribution is recovered at the server. If the updated global imbalance is greater than $\nabla$, then the server exits the proposed protocol, and initiates the FL algorithm among all the clients. On the other hand, if the updated global imbalance is less than $\nabla$, the server performs the following steps:\\
1) If the previous dominant client has not reached its maximum local imbalance, then the same client is chosen as the dominant client even in the current round.\\
2) On the other hand, if the previous client has reached its maximum local imbalance, then the server implements the steps in Section \ref{sec:client_loc} to choose the dominant client by excluding the saturated client(s) in the previous round. Subsequently, the chosen dominant client performs the resampling procedure as explained in the preceding section.

FLICKER terminates either when the desired value of global imbalance is reached, or when all the participating clients have saturated their maximum local imbalance values. 

\subsection{Computational Complexity of FLICKER}

In this section, we characterize the worst-case computational complexity of our method by enumerating the number of times the CKKS based encryption and decryption routines are executed. Recall that CKKS based encryption operation is performed at all the clients when they forward their local distributions during their communication in the ring-based network. As a consequence, the server applies the CKKS decryption operation for recovering the global distribution. Thus, our method requires $N+1$ CKKS operations to gather the initial global distribution.

Furthermore, CKKS encryption is performed on the normalized global distribution at the server in order to identify the dominant client. To assist this task, CKKS encryption is also performed at all the clients for computing the cosine similarities on the ciphertext. Finally, CKKS decryption is employed on the ciphertext received from all the clients. Thus, the total number of CKKS operations to identify the dominant client is $2N + 1$.

After the update process on the local distribution, the dominant client performs CKKS encryption to update the local distribution on the ciphertext of the previous global distribution. Given that there can be at most $M$ rounds of update from a dominant client, for some integer $M$, the number of CKKS encryption and decryption operations is $2M$. Finally, recall that in the worst-case the proposed method terminates after all the clients report saturation on their local imbalance threshold. As a result, assuming that all the clients play the role of dominant client at different points in time, the total number of encryption and decryption operations is at most $2NM$. Overall, the worst-case number of CKKS based operations is $2NM + 3N + 2.$

\section{Resampling Procedures}
\label{RP}

In our resampling method, either the minority class of the dominant client is over-sampled by adding more data samples or the majority class is under-sampled by removing redundant data samples. Given that over-sampling is typically not a favoured option, we implement the over-sampling and the under-sampling methods in an alternate manner until the local imbalance threshold of $\nabla_{local}$ is met.  

\subsection{Over-sampling Procedure}
\label{sec:overlsample}

To perform over-sampling, first, the dominant client identifies the minority class. Subsequently, additional data samples are generated to this class by changing the orientations of the existing samples in this class in case of image datasets. For instance, if the dominant client is client $n$, and its local balance metric is $LI_{n}$, then the client decides to add $\lceil \frac{1}{LI_{n}}\rfloor$ number of additional samples to the minority class. In this context, $\lceil x \rfloor$ denotes the nearest integer to the real number $x$. While carrying out the over-sampling method, we assume that the existing samples are uniformly picked for changing their orientations.

\subsection{Under-sampling Procedure}
\label{sec:under_sampling}

A straightforward way for under-sampling is to randomly remove certain data samples from the majority class. However, it is well known that this approach can lead to loss of important information, which may in turn lead to degraded accuracy numbers. To avoid this information loss, we propose a method, which has two parts, the first part is to identify local redundancy and the second part is to identify global redundancy. 

\begin{table*}
\centering
\begin{tabular}{ |p{1.2cm}|p{2cm}|p{2cm}|p{2cm}|p{2cm}|p{2.5cm}|}
 \hline
 \multicolumn{1}{|c|}{\textit{}} & \multicolumn{1}{c|}{\textit{Class 0}}  & \multicolumn{1}{c|}{\textit{Class 1}} & \multicolumn{1}{c|}{\textit{Class 2}}  & \multicolumn{1}{c|}{\textit{Class 3}} & \multicolumn{1}{c|}{\textit{LI}}\\
 \hline
\textit{Cl-1} & 10/10/2 & 500/30/100 & 700/700/600 & 4000/4000/4000 & 0.0025/0.0025/0.0005\\
 \hline
 \textit{Cl-2} & 20/20/3 & 700/40/200 & 500/500/700 & 3000/3000/3000 & 0.0067/0.0067/0.001\\
 \hline
\textit{Cl-3} & 30/30/5 & 40/40/150 & 600/600/800 & 3000/3000/3000 & 0.01/0.01/0.0017\\
 \hline
 \textit{Cl-4} & 100/50/30 & 50/50/50 & 200/200/20 & 10/10/10 & 0.05/0.05/0.2\\
\hline
\end{tabular}
\caption{(Dataset D1): CIFAR data distribution among four clients capturing one minority class and one majority class. (Dataset D2): CIFAR data distribution among four clients capturing two minority classes and one majority class. (Dataset D3): CIFAR data distribution among four clients capturing high local imbalance among all the clients. Sample numbers are presented as D1/D2/D3 for the three distributions.}
\label{Distribution_1}
\end{table*}

\subsubsection{On Identifying Local Redundancies}
\label{sec:local_red}

We assume that all the clients are equipped with a pre-shared neural network model that has been designed for $K$-class classification task, for some $K >0$. The dominant client feeds its data samples from the majority class and obtains $K$-dimensional feature vectors as a one-to-one correspondence between the data samples and the feature vectors. Subsequently, the feature vectors are normalized to have unit norm, and then the cosine similarities between all the possible pairs of extracted feature vectors are computed. Since these computations are executed locally within the dominant client, cosine similarities are computed in plain-texts in this context. After this step, if the size of the dominant class is $l_{max}$, then a $l_{max} \times l_{max}$ cosine similarity matrix is generated. From this matrix, a certain fraction, namely $\theta \%$ of the feature vectors are picked that have the maximum mean in its cosine similarity values with respect to the other vectors. We also order these $\theta \%$ of the feature vectors with respect to the variance of cosine similarities. The variance metric is significant since the samples with higher variance indicates that they have higher similarity with more number of data samples in the class. These feature vectors are buffered in order to use them to compute global redundancy among the samples with the other clients.  

\subsubsection{On Identifying Global Redundancies}
\label{sec:global_red}

To verify global redundancy of the selected samples, a buffered feature vector is encrypted using CKKS and is sent to the server. Then the server forwards it to the rest of the clients where every client computes the cosine similarity with its feature vectors on the cipher-text. Finally, after the decryption operation at the server, the cosine similarities of each client are forwarded to the dominant client through the server. 
For a given buffered vector, if $\chi \%$ of the global feature vectors, for some $\chi$, have cosine similarities above a certain threshold, say $\eta$, for some $1 < \eta < 1$, then the buffered training sample is removed by the client. The same procedure is repeated for all the the buffered feature vectors to complete the under-sampling procedure. 

\section{Experimental Results}

To present the experimental results, we first describe the settings used for conducting the experiments, then present the baselines, and finally discuss the merits of implementing FLCIKER.

\subsection{Experimental Settings}

Our FL setup consists of four clients and one server, wherein the clients are interested in executing a classification task among four classes. For implementation purposes, the training and test data samples are carved from the first four classes of CIFAR-10, and the clients use a custom-built convolutional neural network (CNN) for classification. In particular, for the CIFAR-10 dataset, the CNN classifier at each client uses 1,288,256 weight parameters, 16,490 bias parameters, with a total of 1,304,746 parameters. To provide diverse set of experiments to study class imbalance issues, we consider the initial data distributions as listed in Table \ref{Distribution_1}. It can be observed that various degrees of imbalances and minority/majority classes can be witnessed through these distributions. The corresponding local balance metrics are also reported in the same tables.

To evaluate the efficacy of FLICKER on imbalanced datasets from diverse domains, we also implement our strategy on AG news datasets, wherein the task is to classify a news article into one of the following four classes, namely: (i) world, (ii) sports, (iii) business, and (iv) science/tech, based on the headlines and a short description of the news. When using the AG news datasets, the data distribution shown in Table \ref{Distribution_ag_news} is used. Similar to the CIFAR dataset, the clients are assumed to use a custom-built CNN in the NLP context for classification. In particular, the CNN classifier for AG news at each client uses $644,352$ weight parameters, $68$ bias parameters, with a total of $644420$ parameters.

\begin{table*}
\centering
\begin{tabular}{ |p{1.2cm}|p{2cm}|p{2cm}|p{2cm}|p{2cm}|p{2cm}|}
 \hline
 \multicolumn{1}{|c|}{\textit{}} & \multicolumn{1}{c|}{\textit{Class 0}}  & \multicolumn{1}{c|}{\textit{Class 1}} & \multicolumn{1}{c|}{\textit{Class 2}}  & \multicolumn{1}{c|}{\textit{Class 3}} & \multicolumn{1}{c|}{\textit{LI}}\\
\hline
\textit{Precision} & 0.91/0.84/0.87 & 0.73/0.88/0.79 & 0.60/0.47/0.51 & 0.59/0.45/0.52 & -\\
 \hline
 \textit{Recall} & 0.31/0.28/0.14 & 0.90/0.35/0.75 & 0.55/0.63/0.60 & 0.88/0.88/0.90 & -\\
 \hline
\textit{F1-score} & 0.46/0.42/0.24 & 0.81/0.5/0.77 & 0.57/0.54/0.55 & 0.70/0.60/0.66 & -\\
 \hline
 \multicolumn{6}{|c|}{\textit{Global Imbalance = 0.016/0.016/0.0039}}\\
 \hline
 \multicolumn{6}{|c|}{\textit{Final Accuracy = 66.04\%/53.45\%/59.77\%}}\\
 \hline
\end{tabular}
\caption{Results on FL when the initial distributions of CIFAR datasets in Table \ref{Distribution_1}, are used along with no imbalance correction.}
\label{dist_1_no_correction}
\end{table*}


\begin{table*}
\centering
\begin{tabular}{ |p{1.5cm}|p{2cm}|p{2cm}|p{2cm}|p{2cm}|p{2cm}|}
 \hline
 \multicolumn{1}{|c|}{\textit{}} & \multicolumn{1}{c|}{\textit{Class 0}}  & \multicolumn{1}{c|}{\textit{Class 1}} & \multicolumn{1}{c|}{\textit{Class 2}}  & \multicolumn{1}{c|}{\textit{Class 3}} & \multicolumn{1}{c|}{\textit{LI}}\\
 \hline
\textit{Cl-1} & 10/10/2 & 155/30/40 & 170/171/29 & 146/112/28 & 0.058/0.058/0.5\\
 \hline
 \textit{Cl-2} & 20/30/3 & 373/40/60 & 400/392/57 & 368/364/56 & 0.05/0.051/0.05\\
 \hline
\textit{Cl-3} & 30/30/3 & 400/40/85 & 600/600/92 & 500/566/99 & 0.05/0.05/0.03\\
 \hline
 \textit{Cl-4} & 100/50/30 & 50/50/50 & 200/200/20 & 10/10/10 & 0.05/0.05/0.2\\
 \hline
\textit{Precision} & 0.90/0.81/1 & 0.74/0.84/0.53 & 0.46/0.42/0.39 & 0.51/0.33/0.39 & -\\
 \hline
 \textit{Recall} & 0.28/0.20/0.01 & 0.86/0.38/0.77 & 0.80/0.82/0.57 & 0.41/0.44/0.41 & -\\
 \hline
\textit{F1-score} & 0.43/0.32/0.01 & 0.79/0.52/0.63 & 0.59/0.56/0.46 & 0.46/0.38/0.4 & -\\
 \hline
 \multicolumn{6}{|c|}{\textit{Samples Removed = 10288/9395/12004}}\\
 \hline
 \multicolumn{6}{|c|}{\textit{Global Imbalance = 0.117/0.081/0.1617}}\\
 \hline
 \multicolumn{6}{|c|}{\textit{Final Accuracy = 58.77\%/46.00\%/44.67\%}}\\
 \hline
\end{tabular}
\caption{Results on FL when the initial distributions of CIFAR datasets in Table \ref{Distribution_1} are corrected using under-sampling.}
\label{dist_1_under_sampling_correction}
\end{table*}

\begin{table*}
\centering
\begin{tabular}{ |p{1.2cm}|p{2cm}|p{2cm}|p{2cm}|p{2cm}|p{2.5cm}|}
 \hline
 \multicolumn{1}{|c|}{\textit{}} & \multicolumn{1}{c|}{\textit{Class 0}}  & \multicolumn{1}{c|}{\textit{Class 1}} & \multicolumn{1}{c|}{\textit{Class 2}}  & \multicolumn{1}{c|}{\textit{Class 3}} & \multicolumn{1}{c|}{\textit{LI}}\\
 \hline
\textit{Cl-1} & 410/410/1894 & 500/580/2092 & 700/700/600 & 4000/4000/4000 & 0.1025/0.1025/0.2868\\
 \hline
 \textit{Cl-2} & 180/180 & 700/280/200 & 500/500 & 3000/3000/3000 & 0.06/0.06/0.667\\
 \hline
\textit{Cl-3} & 150/150/605 & 400/240/150 & 600/600/800 & 3000/3000/3000 & 0.05/0.05/0.05\\
 \hline
 \textit{Cl-4} & 100/50/30 & 50/50/50 & 200/200/20 & 10/10/10 & 0.05/0.05/0.2\\
 \hline
\textit{Precision} & 0.73/0.75/0.69 & 0.81/0.78/0.76 & 0.66/0.55/0.54 & 0.60/0.49/0.54 & -\\
 \hline
 \textit{Recall} & 0.53/0.38/0.28 & 0.82/0.55/0.73 & 0.55/0.53/0.51 & 0.86/0.89/0.91 & -\\
 \hline
\textit{F1-score} & 0.62/0.5/0.4 & 0.82/0.65/0.74 & 0.60/0.54/0.53 & 0.71/0.63/0.67 & -\\
 \hline
 \multicolumn{6}{|c|}{\textit{Samples Augmented = 680/1670/5465}}\\
 \hline
 \multicolumn{6}{|c|}{\textit{Global Imbalance = 0.084/0.078/0.2107}}\\
 \hline
 \multicolumn{6}{|c|}{\textit{Final Accuracy = 69.07\%/58.85\%/59.77\%}}\\
 \hline
\end{tabular}
\caption{Results on FL when the initial distributions of CIFAR datasets in Table \ref{Distribution_1} are corrected using over-sampling.}
\label{dist_1_over_sampling_correction}
\end{table*}

\begin{table*}
\centering
\begin{tabular}{ |p{1.2cm}|p{2cm}|p{2cm}|p{2cm}|p{2cm}|p{3cm}|}
 \hline
 \multicolumn{1}{|c|}{\textit{}} & \multicolumn{1}{c|}{\textit{Class 0}}  & \multicolumn{1}{c|}{\textit{Class 1}} & \multicolumn{1}{c|}{\textit{Class 2}}  & \multicolumn{1}{c|}{\textit{Class 3}} & \multicolumn{1}{c|}{\textit{LI}}\\
 \hline
 \textit{Cl-1} & 398/400/493 & 500/360/200 & 700/700/600 & 3849/3840/3968 & 0.103/0.1041/0.05\\
 \hline
 \textit{Cl-2} & 180/180/206 & 700/240/200 & 500/500/700 & 2830/2909/2455 & 0.064/0.0618/0.0814\\
 \hline
 \textit{Cl-3} & 240/120/130 & 400/120/150 & 600/600/800 & 2538/2304/2490 & 0.095/0.0520/0.0522\\
 \hline
 \textit{Cl-4} & 100/50/30 & 50/50/50 & 200/200/20 & 10/10/10 & 0.05/0.05/0.2\\
 \hline
\textit{Precision} & 0.77/0.77/0.76 & 0.81/0.72/0.80 & 0.65/0.58/0.55 & 0.59/50/0.54 & -\\
 \hline
 \textit{Recall} & 0.50/0.32/0.32 & 0.83/0.62/0.77 & 0.54/0.52/0.56 & 0.87/0.88/0.87 & -\\
 \hline
\textit{F1-score} & 0.61/0.45/0.45 & 0.82/0.69/0.79 & 0.59/0.55/0.56 & 0.70/0.64/0.67 & -\\
 \hline
 \multicolumn{6}{|c|}{\textit{Overall Change in Data Samples Count = -25/303/-168}}\\
 \hline
 \multicolumn{6}{|c|}{\textit{Global Imbalance = 0.0994/0.083/0.0672}}\\
 \hline
 \multicolumn{6}{|c|}{\textit{Final Accuracy = 68.73\%/59.60\%/63.05\%}}\\
 \hline
\end{tabular}
\caption{Results on FL when the initial distributions of CIFAR datasets in Table \ref{Distribution_1} are corrected corrected using FLICKER.}
\label{dist_1_our_method_correction}
\end{table*}

To implement the CKKS protocol in Section \ref{sec:sgic} and Section \ref{sec:client_loc}, the threshold required on the global imbalance at the server is $\nabla = 0.1$. On the other hand, threshold on the local imbalance at each client is  $\nabla_{client} = 0.05$. To implement the under-sampling method in Section \ref{sec:under_sampling}, all the clients need a pre-shared model that has been trained to execute classification tasks among $K$ classes. For this, the last 1000 data samples of the training dataset from all four classes are assumed to be auxiliary data. Furthermore, a separate classification model is trained with this auxiliary data for 50 epochs, and the instances of this model were shared with all the clients for extraction of feature vectors. Therefore $K = 4$ in this case. Finally, to implement the local and global redundancy tests in Section \ref{sec:under_sampling}, we use $\theta = 10$, $\eta = 0.98$.

After the proposed CKKS method is implemented on the CIFAR and AG news datasets, we implement the FL operation to obtain the global model, and then test the accuracy of the global model after each round for a total of 20 rounds. Similar set of experiments are repeated wherein accuracy curves are plotted for the global model trained with original imbalanced dataset without implementing any imbalance correction methods. As the main departure from the traditional definition of accuracy in FL settings, we introduce the notion of \emph{normalized accuracy} in our experimental results, which is defined to accommodate the space-overheads resulting from the over-sampling methods for correcting dataset imbalance. For an FL setup with the underlying ML models designed for a classification task, let $X = \sum_{n=1}^N\sum_{j = 1}^L l_{n,j}$ be the total number of training samples across all the classes and all the clients. For the same setup, if the clients undergo a resampling procedure to correct the imbalance in their datasets, let $Y$ be the number of total number of training samples across all the classes and all the clients, and let $A$ be the test accuracy of the trained FL model after imbalance correction. In such scenarios, normalized accuracy, denoted by $A_{N}$ is 
\begin{align}
\label{norm_accuracy}
A_{N} = \frac{A}{max(1, \frac{Y}{X})}.
\end{align}
Note that \eqref{norm_accuracy} penalizes any accuracy benefits offered by over-sampling methods due to additional \emph{artificial} data samples inserted in the training dataset. On the other hand, \eqref{norm_accuracy} does not alter the accuracy of under-sampling methods as they do not introduce any additional space-overheads at the clients. Thus, normalized accuracy serves as an excellent metric to compare the efficacy of various resampling methods when implemented on space-constrained devices, often witnessed in Internet of Things use-cases \cite{IOTSurvey}.

\subsection{Baselines}

For a comparative analysis of FLICKER, we use the following two baselines, wherein clients individually correct their local imbalance without checking the global imbalance. 

1) \textit{Under-sampling baseline}: In this setup, all the clients correct their own local imbalance by removing samples from their majority class. To remove the redundancy in their data samples, we assume that the clients use the method in Section \ref{sec:local_red}, wherein only the local redundancy in the majority class is removed with $\theta = 10$. Similar to our method, the clients thrive to achieve a local imbalance threshold of $\nabla_{client} = 0.05$.

2) \textit{Over-sampling baseline}: In this setup, all the clients correct their own local imbalance by augmenting data samples in their minority class. This over-sampling procedure is the same as in Section \ref{sec:overlsample}. 

\subsection{Results}

In this section, we present the benefits of FLICKER over the baselines by focusing on the distributions in Table \ref{Distribution_1} for the CIFAR dataset and by focusing on the distribution in Table \ref{Distribution_ag_news} for the AG news dataset.. 

\subsubsection{Results on the Distributions from CIFAR-10 Dataset}

\begin{figure}[htpb]
  \begin{center}
   \includegraphics[scale = 0.35]{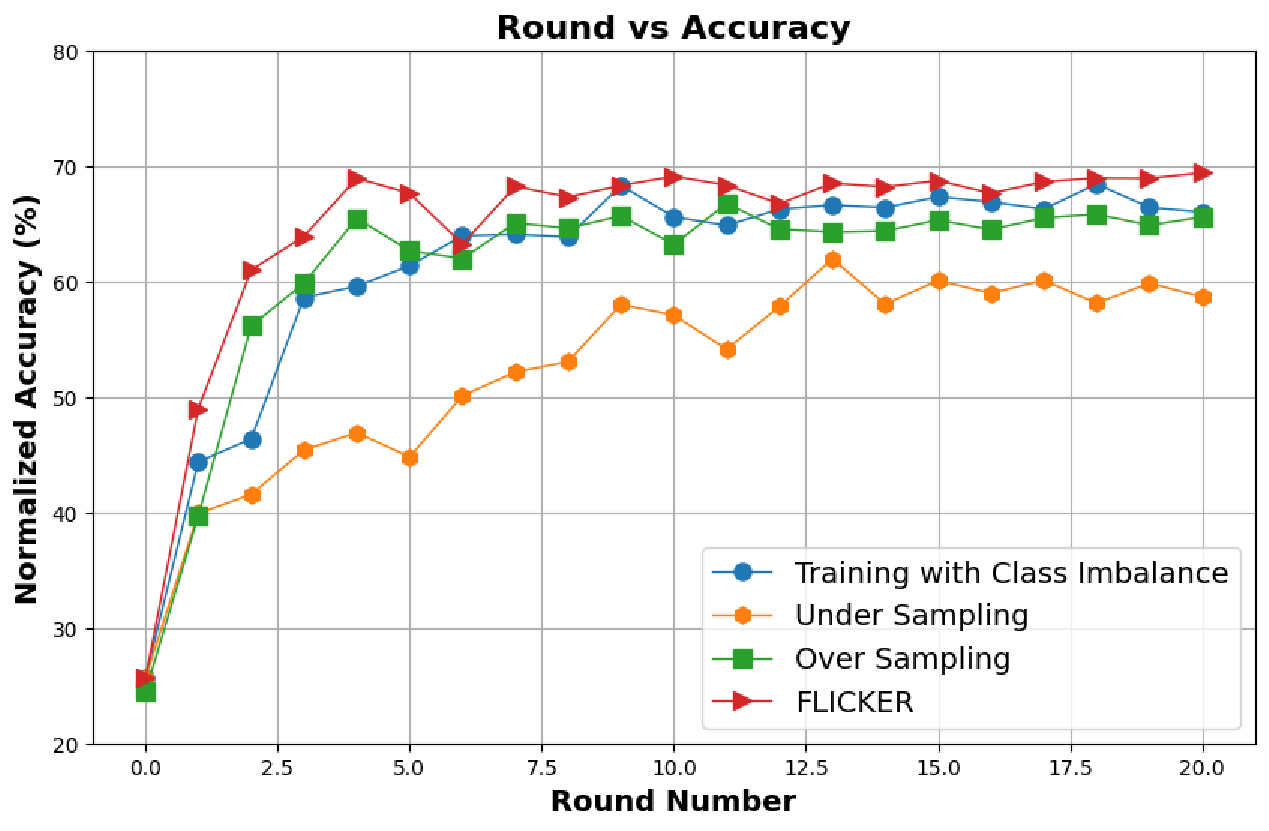}  
   \vspace{-0.5cm}
    \caption{FL accuracy comparison as a function of the training rounds when the initial distribution D1 of CIFAR datasets.}
    \label{FL_dist_1}
  \end{center}
\end{figure}

For the CIFAR dataset with class-wise sample distribution in Table~\ref{Distribution_1}, we trained our global model with the imbalance data distributions without correction. The corresponding overall accuracy numbers after 20 rounds of training, and the values of F1-scores after twenty rounds of training for respective classes are listed in Table~\ref{dist_1_no_correction}. Thereafter, we applied the under-sampling and the over-sampling baseline methods for the clients to individually correct their respective local imbalances. Details about their new distribution of data samples, new global Imbalance along with their F1-scores and model accuracy after 20 rounds are given in  Table~\ref{dist_1_under_sampling_correction} and Table~\ref{dist_1_over_sampling_correction}, respectively. Finally, we applied FLICKER to detect the global class imbalance and carryout resampling at clients while preserving dataset privacy of respective clients. Details about the updated distribution of data samples, updated global imbalance, F1-scores, and model accuracy after 20 rounds are given in Table~\ref{dist_1_our_method_correction}. The corresponding normalized accuracy curves, as defined in \eqref{norm_accuracy}, comparing our method with the baselines are as given in Fig. \ref{FL_dist_1}. As it can be seen from tables and the accuracy curves, there is a prominent increase in the normalized accuracy and the F1-scores post applying FLICKER. While the under-sampling baseline performs poorly, the over-sampling baseline is marginally below our FLICKER method. Note that this difference is because the over-sampling baseline augments 680 additional samples to the local distributions which is not desired. 

\begin{table*}
\centering
\begin{tabular}{ |p{1.2cm}|p{2.5cm}|p{2.5cm}|p{2.5cm}|p{2.5cm}|p{3.5cm}|}
 \hline
 \multicolumn{1}{|c|}{\textit{}} & \multicolumn{1}{c|}{\textit{Class 0}}  & \multicolumn{1}{c|}{\textit{Class 1}} & \multicolumn{1}{c|}{\textit{Class 2}}  & \multicolumn{1}{c|}{\textit{Class 3}} & \multicolumn{1}{c|}{\textit{LI}}\\
 \hline
 \textit{Cl-1} & 10/10/410/389 & 500/140/500/500 & 700/177/700/700 & 4000/191/4000/3793 & 0.0025/0.0523/0.1025/0.1026\\
 \hline
 \textit{Cl-2} & 20/20/170/161 & 700/373/700/700 & 500/391/500/500 & 3000/268/3000/2824 & 0.0067/0.0522/0.0567/0.0570\\
 \hline
 \textit{Cl-3} & 30/30/183/144 & 400/400/400/400 & 600/600/600/600 & 3000/456/3000/2633 & 0.01/0.05/0.061/0.0547\\
 \hline
 \textit{Cl-4} & 100/100/100/100 & 50/50/50/50 & 200/200/200/200 & 10/10/10/10 & 0.05/0.05/0.05/0.05\\
 \hline
\textit{Precision} & 0.90/0.90/0.97/0.97 & 0.89/0.89/0.90/0.90 & 0.75/0.55/0.70/0.71 & 0.68/0.52/0.71/0.71 & -\\
 \hline
 \textit{Recall} & 0.46/0.33/0.46/0.48 & 0.95/0.94/0,96/0.96 & 0.85/0.93/0.87/0.88 & 0.87/0.47/0.86/0.86 & -\\
 \hline
\textit{F1-score} & 0.61/0.48/0.63/0.64 & 0.92/0.92/0.93/0.93 & 0.79/0.69/0.78/0.78 & 0.76/0.50/0.78/0.78 & -\\
 \hline
 \multicolumn{6}{|c|}{\textit{Overall Change in Data Samples Count = 0/-10404/703/-116}}\\
 \hline
 \multicolumn{6}{|c|}{\textit{Global Imbalance = 0.016/0.1170/0.0862/0.0458}}\\
 \hline
 \multicolumn{6}{|c|}{\textit{Final Accuracy = 78.38\%/66.68\%/74.96\%/79.26\%}}\\
 \hline
\end{tabular}
\caption{Results on FL using with-imbalance/under-sampling/over-sampling/FLICKER on AG news text classification dataset.}
\label{Distribution_ag_news}
\end{table*}

\subsubsection{Results on the Distribution from AG news datasets}

\begin{figure}[ht!]
  \begin{center}
   \includegraphics[scale = 0.35]{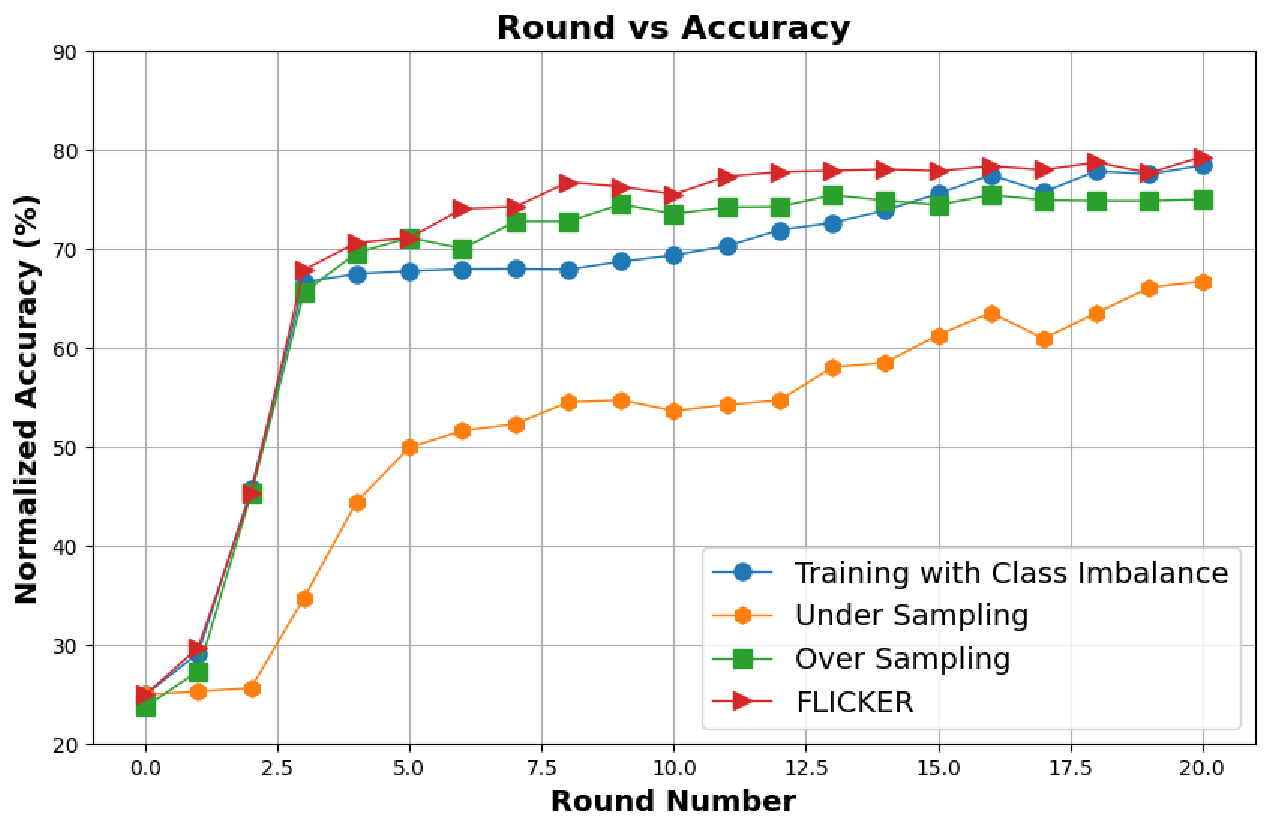}
   \vspace{-0.5cm}
    \caption{FL accuracy comparison as a function of the training rounds when the initial distribution of AG news datasets in Table \ref{Distribution_ag_news} is used.}
    \label{normalized_accurcy_AG_news}
  \end{center}
\end{figure}

When the four clients apply FL on these datasets without any imbalance correction, they achieve the accuracy numbers and the F1-scores as listed in the first set of numbers in Table \ref{Distribution_ag_news}. However, after applying FLICKER, and subsequent FL operation, the clients achieve accuracy numbers and the F1-scores as listed in the fourth set of numbers in the same table. Finally, the normalized accuracy numbers in Fig. \ref{normalized_accurcy_AG_news} confirms that FLICKER outperforms the baselines along the similar lines of the CIFAR datasets. 

\begin{figure}[ht!]
  \begin{center}
   \includegraphics[scale = 0.35]{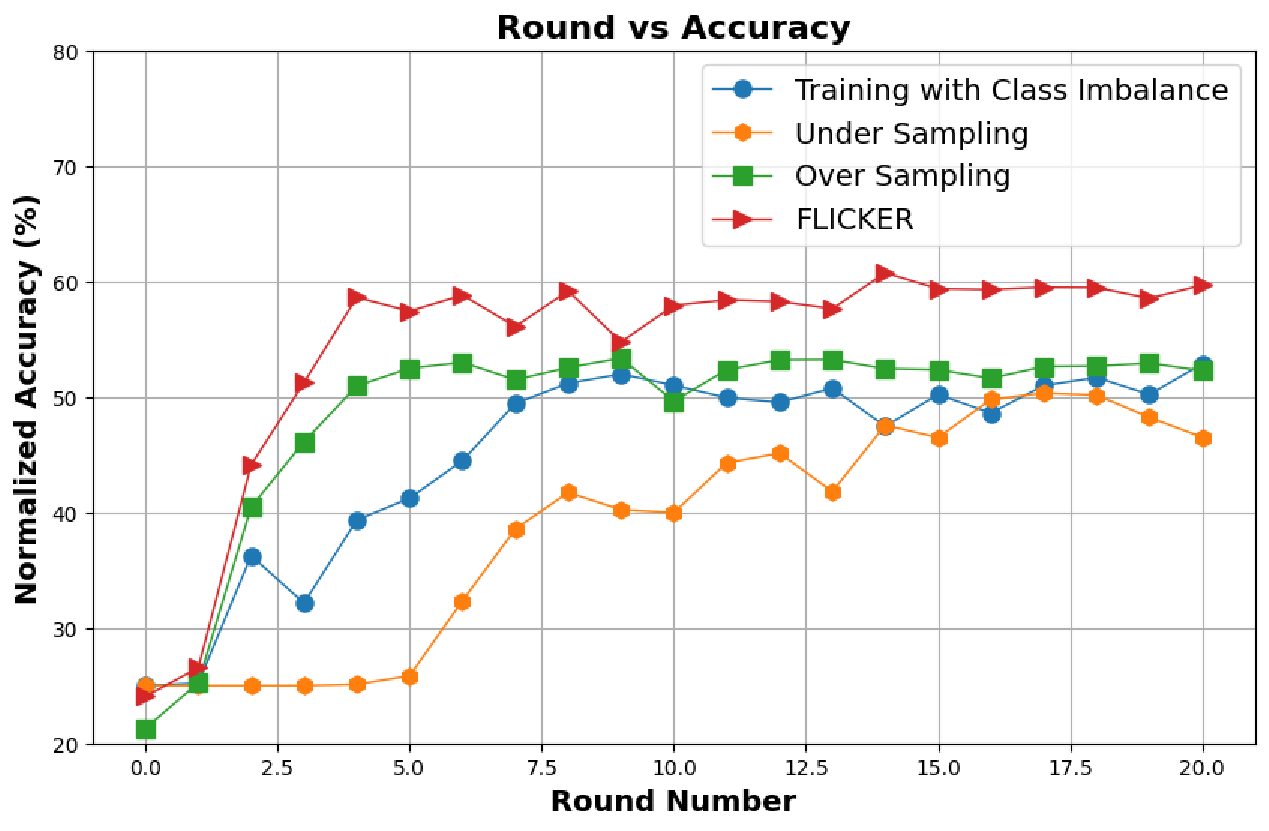}
   \vspace{-0.5cm}
    \caption{FL accuracy comparison as a function of the training rounds when the initial distribution D2 of CIFAR datasets.}
    \label{FL_dist_2}
  \end{center}
\end{figure}

\begin{figure}[ht!]
  \begin{center}
   \includegraphics[scale = 0.35]{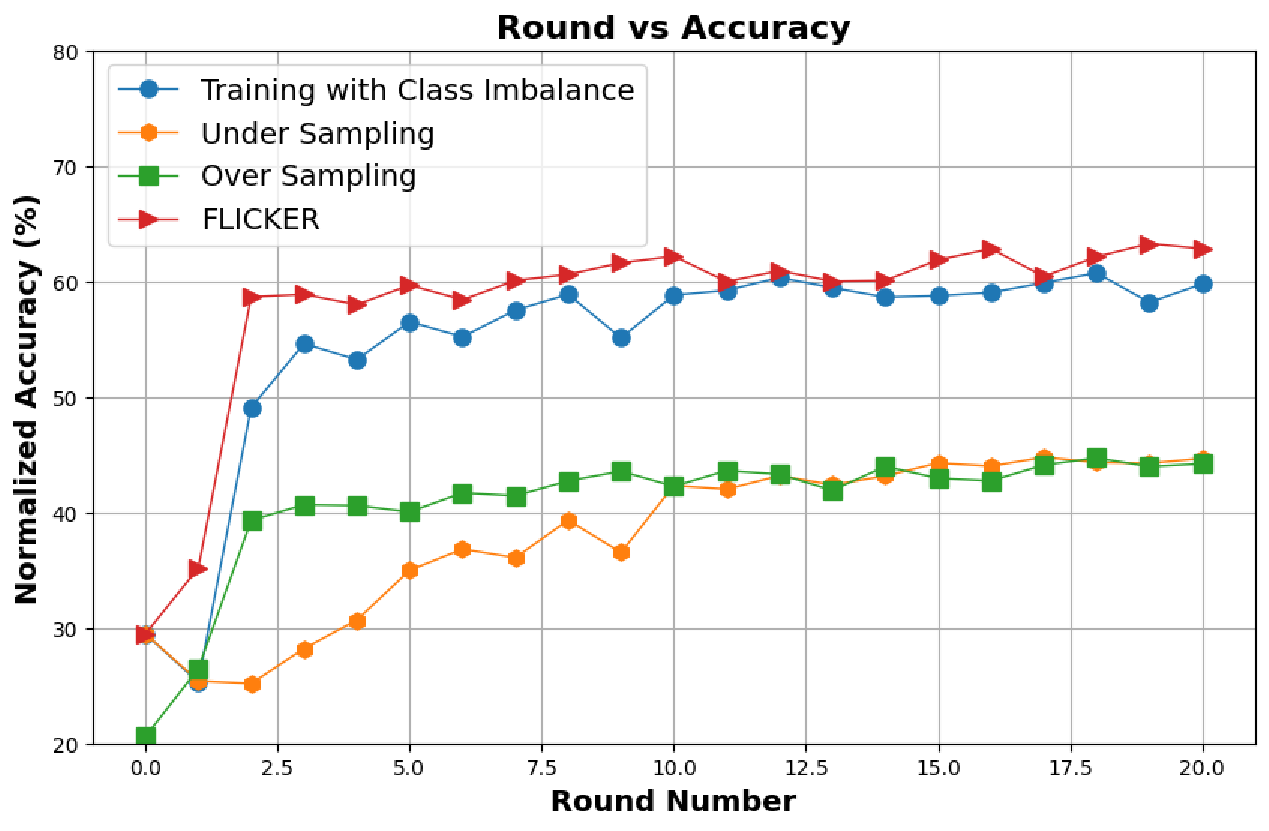}
   \vspace{-0.5cm}
    \caption{FL accuracy comparison as a function of the training rounds when the initial distribution D3 of CIFAR datasets.}
    \label{FL_dist_4}
  \end{center}
\end{figure}

 \section{Directions for Future Research}

For the FL framework, we have suggested a way to detect and correct the global imbalance in the global distribution of data samples using the CKKS homomorphic scheme. Since only vector addition operation is performed for detection, computational-overheads are not overwhelming. However, in general, a major issue in using homomorphic encryption schemes is overheads involved due to encryption and homomorphic operations. Given that one of the major applications of FL is among wireless mobile devices, light-weight homomorphic schemes \cite{LW-1}, \cite{LW-2}, \cite{LW-3} could be promising solutions to reduce the overhead problems while still preserving the data privacy efficiently.


\begin{thebibliography}{16}
\bibliographystyle{IEEEtran}

\bibitem{FLApplSurvey}
Wen, Jie, et al. "A survey on federated learning: challenges and applications." International Journal of Machine Learning and Cybernetics 14.2 (2023): 513-535.

\bibitem{Fedavg}
McMahan, H. Brendan and Moore, Eider and Ramage, Daniel and Hampson, Seth and Arcas, Blaise Agüera,"Communication-{Efficient} {Learning} of {Deep} {Networks} from {Decentralized} {Data}", AISTATS 2017

\bibitem{FL_Homomorphic}
Park, Jaehyoung, and Hyuk Lim. "Privacy-preserving federated learning using homomorphic encryption." Applied Sciences 12.2 (2022): 734.

\bibitem{Glo_Imbal}
C. Xiao and S. Wang, ``An Experimental Study of Class Imbalance in Federated Learning," in \emph{2021 IEEE Symposium Series on Computational Intelligence (SSCI), Orlando, FL, USA, 2021,} pp. 1-7.

\bibitem{LearnImbData}
He, H., and Garcia, E. A. (2009). Learning from imbalanced data. IEEE Transactions on knowledge and data engineering, 21(9), 1263-1284.



\bibitem{datasamp}
Cui, Y., Jia, M., Lin, T. Y., Song, Y., and Belongie, S. (2019). Class-balanced loss based on effective number of samples. In Proceedings of the IEEE/CVF conference on computer vision and pattern recognition (pp. 9268-9277).

\bibitem{ADASYN}
He, H., Bai, Y., Garcia, E. A., and Li, S., "ADASYN: Adaptive synthetic sampling approach for imbalanced learning", \emph{IEEE international joint conference on neural networks}, pp. 1322-1328.

\bibitem{IOTSurvey}
 C. Sabri, L. Kriaa, and S. L. Azzouz. 2017. "Comparison of IoT Constrained Devices Operating Systems: A Survey."  \emph{IEEE/ACS 14th International Conference on Computer Systems and Applications (AICCSA)}

\bibitem{HEAN}
H. J. Cheon, A. Kim, M. Kim and Y. Song, ``Homomorphic Encryption for Arithmetic of Approximate Numbers" in 2017 Advances in Cryptology-ASIACRYPT 2017. ASIACRYRT 2017. Lecture Notes in Computer Science, Springer, November 2017.

\bibitem{CIF1}
Wang L Xu, S., Wang X., and Zhu Q. (2021). Addressing Class Imbalance in Federated Learning. Proceedings of the AAAI Conference on Artificial Intelligence, 35(11), 10165-10173. 

\bibitem{CIF2}
M. Yang, X. Wang, H. Zhu, H. Wang and H. Qian, "Federated Learning with Class Imbalance Reduction," \emph{2021 29th European Signal Processing Conference (EUSIPCO), Dublin, Ireland, 2021}, pp. 2174-2178,

\bibitem{CIF3}
Zebang Shen and Juan Cervino and Hamed Hassani and Alejandro Ribeiro, ``An Agnostic Approach to Federated Learning with Class Imbalance " in \emph{shen2022an,International Conference on Learning Representations 2022}

\bibitem{MID}
C. Xiao and S. Wang, "An Experimental Study of Class Imbalance in Federated Learning," 2021 IEEE Symposium Series on Computational Intelligence (SSCI), Orlando, FL, USA, 2021, pp. 1-7


\bibitem{longtailed}
X. Shuai, Y. Shen, S. Jiang, Z. Zhao, Z. Yan and G. Xing, "BalanceFL: Addressing Class Imbalance in Long-Tail Federated Learning," ACM/IEEE International Conference on Information Processing in Sensor Networks (IPSN) 2022, pp. 271-284

\bibitem{surveyimbalance}
Zhang, J., Li, C., Qi, J., and He, J. (2023). A survey on class imbalance in federated learning, arXiv preprint arXiv:2303.11673.

\bibitem{CIClientSelec}
Yang, M., Qian, H., Wang, X., Zhou, Y., and Zhu, H. (2021). Client selection for federated learning with label noise. IEEE Transactions on Vehicular Technology, 71(2), 2193-2197.

\bibitem{Ijcai23}
Wu, Nannan and Yu, Li and Jiang, Xuefeng and Cheng, Kwang-Ting and Yan, Zengqiang. FedNoRo: Towards Noise-Robust Federated Learning by Addressing Class Imbalance and Label Noise Heterogeneity, IJCAI 2023


\bibitem{CKKS_FL_1}
Pan, Y., Chao, Z., He, W. et al., FedSHE: privacy preserving and efficient federated learning with adaptive segmented CKKS homomorphic encryption, \emph{Cybersecurity} 7, 40 (2024).

\bibitem{CKKS_FL_2}
Qiu, Fengyuan, et al. ``Privacy preserving federated learning using ckks homomorphic encryption." \emph{International Conference on Wireless Algorithms, Systems, and Applications}, Cham: Springer Nature Switzerland, 2022.

\bibitem{CKKS_FL_3}
Ma, Jing, et al. ``Privacy‐preserving federated learning based on multi‐key homomorphic encryption," \emph{International Journal of Intelligent Systems} 37.9 (2022): 5880-5901.

\bibitem{CKKS_FL_4}
Stripelis, Dimitris, et al. ``Secure neuroimaging analysis using federated learning with homomorphic encryption," \emph{17th International Symposium on Medical Information Processing and Analysis}, Vol. 12088. SPIE, 2021.

\bibitem{FE_FL_1}
N. M. Hijazi, M. Aloqaily, M. Guizani, B. Ouni and F. Karray, ``Secure Federated Learning With Fully Homomorphic Encryption for IoT Communications," in \emph{IEEE Internet of Things Journal}, vol. 11, no. 3, pp. 4289-4300, 1 Feb.1, 2024,

\bibitem{LW-1}
M. R. Baharon, Q. Shi and D. Llewellyn-Jones, ``A New Lightweight Homomorphic Encryption Scheme for Mobile Cloud Computing," 2015 \emph{IEEE International Conference on Computer and Information Technology; Ubiquitous Computing and Communications; Dependable, Autonomic and Secure Computing; Pervasive Intelligence and Computing, Liverpool, UK, 2015,} pp. 618-625, doi: 10.1109/CIT/IUCC/DASC/PICOM.2015.88.

\bibitem{LW-2}
S. Li, S. Zhao, G. Min, L. Qi and G. Liu, ``Lightweight Privacy-Preserving Scheme Using Homomorphic Encryption in Industrial Internet of Things," in \emph{IEEE Internet of Things Journal, vol. 9, no. 16, pp. 14542-14550,} 15 Aug.15, 2022,

\bibitem{LW-3}
Abdulatif Alabdulatif, Heshan Kumarage, Ibrahim Khalil, Xun Yi, ``Privacy-preserving anomaly detection in cloud with lightweight homomorphic encryption," in \emph{Journal of Computer and System Sciences, Volume 90,2017} Pages 28-45,
























\end{thebibliography}
\end{document}